\documentclass[pra,twocolumn,superscriptaddress,showpacs,preprintnumbers,amsmath,amssymb]{revtex4-1}
\makeatletter
\usepackage[dvips]{graphicx}

\usepackage{amsmath}
\usepackage{braket}
\usepackage{amssymb}
\usepackage{amsbsy}
\usepackage{color}
\setcitestyle{square}
\usepackage{epstopdf}
\DeclareGraphicsExtensions{.pdf,.eps,.png,.jpg,.mps}

\definecolor{textcolor}{cmyk}{0,0,0,1}
\definecolor{magenta}{rgb}{1,0,1}
\definecolor{green}{rgb}{0,1,0}
\definecolor{red}{rgb}{1,0,0}

\begin{document}

\title{
Controlling the layer localization of gapless states 
in bilayer graphene with a gate voltage
}
\author{W. Jask\'olski}
\affiliation{Institute of Physics, Faculty of Physics, Astronomy and Informatics, Nicolaus Copernicus University, Grudziadzka 5, 87-100 Toru\'n, Poland}
\author{M. Pelc}
\email{marta\_pelc001@ehu.eus}
\affiliation{Donostia International Physics Center (DIPC), Paseo Manuel Lardizabal 4, 20018 Donostia-San Sebasti\'an, Spain}
\affiliation{Centro de F\'isica de Materiales, CFM-MPC CSIC-UPV/EHU, Paseo Manuel Lardizabal 5, 20018 Donostia-San Sebasti\'an, Spain}
\affiliation{Institute of Physics, Faculty of Physics, Astronomy and Informatics, Nicolaus Copernicus University, Grudziadzka 5, 87-100 Toru\'n, Poland}
\author{Garnett W. Bryant}
\affiliation{Quantum Measurement Division and Joint Quantum Institute, National Institute of Standards and Technology, Gaithersburg, MD, USA 20899-8423}
\author{Leonor Chico}
\affiliation{Instituto de Ciencia de Materiales de Madrid (ICMM), Consejo Superior de Investigaciones Cient\'ificas (CSIC), C/ Sor Juana In\'es de la Cruz 3, 28049 Madrid, Spain}
\affiliation{Donostia International Physics Center (DIPC), Paseo Manuel Lardizabal 4, 20018 Donostia-San Sebasti\'an, Spain}
\author{A. Ayuela}
\email{swxayfea@ehu.eus}
\affiliation{Donostia International Physics Center (DIPC), Paseo Manuel Lardizabal 4, 20018 Donostia-San Sebasti\'an, Spain}
\affiliation{Centro de F\'isica de Materiales, CFM-MPC CSIC-UPV/EHU, Paseo Manuel Lardizabal 5, 20018 Donostia-San Sebasti\'an, Spain}
\affiliation{Departamento de F\'isica de Materiales, Facultad de Qu\'imicas, UPV-EHU, 20018 San Sebasti\'an, Spain}

\date{\today}

%%%%%%%%%%%%%%%%%%%%%%%%%%%%%%%%%%%%%%%%%%%%%%%%%%%%%%%%%%%%%%%%%%%%%%%%%%%%

\begin{abstract} 
Experiments in gated bilayer graphene with stacking domain walls present topological gapless states protected by no-valley mixing. Here we research these states under gate voltages using atomistic models, which allow us to elucidate their origin. We find that the gate potential controls the layer localization of the two states, which switches non-trivially between layers depending on the applied gate voltage magnitude. We also show how these bilayer gapless states arise from bands of single-layer graphene by analyzing the formation of carbon bonds between layers. Based on this analysis we provide a model Hamiltonian with analytical solutions, which explains the layer localization as a function of the ratio between the applied potential and interlayer hopping. Our results open a route for the manipulation of gapless states in electronic devices, analogous to the proposed writing and reading memories in topological insulators. 

\end{abstract}


\maketitle

%%%%%%%%%%%%%%%%%%%%%%%%%%%%%%%%%%%%%%%%%%%%%%%%%%%%%%%%%%%%%%%%%%%%%%%%%%%%%%% INTRODUCTION %%%
%%%%%%%%%%%%%%%%%%%%%%%%%%%%%%%%%%%%%%%%%%%%%%%%%%%%%%%%%%%%%%%%%%%%%%%%%%%
\section{\label{sec:intro} Introduction}

Two dimensional Dirac materials, like graphene, have attracted remarkable interest for novel nanoelectronic applications 
due to their reduced dimensionality and extraordinary transport properties \cite{Novoselov_2005,Neto_2009,Sarma_2011,Chiatti_2016,Cahangirov_2009,Ma_2014}.
Recently, great effort has been devoted to seek and develop methods to control the transport of different degrees of freedom in these new materials. Spin \cite{Han_2014}, valley \cite{Shimazaki_2015,Sui_2015}, angular momentum \cite{Manchon_2015} or cone \cite{Menghao_2017} is being proposed in addition to charge as a means to convey and store information in future devices. In this field, bilayer graphene (BLG) stands out as a promising candidate for nanoelectronics \cite{Castro_2007,Ohta_2006,Szafranek_2010,Zhang_2009,Santos_2012}. In its most common form, the so-called Bernal or AB-stacked BLG, an energy gap can be opened and tuned by an applied gate voltage \cite{Schwierz_2010,Lin_2008,Choi_2010,Padilha_2011}, which is not possible in single-layer graphene.

The gap in AB-stacked BLG arises because of the combination of two factors: the interlayer hopping permits to distinguish A and B atoms, breaking sublattice symmetry, and the gate potential breaks inversion symmetry, differentiating the two layers (Fig. 1 left). Importantly, when a domain wall (DW) divides bilayer graphene into AB and BA stacking domains, a pair of states appears at each valley, connecting the valence and conduction band continua through the energy gap (Fig. 1 right) \cite{Alden_2013,San_Jose_2014,Pelc_2015}. These gapless states are topologically protected if the valley index is conserved. Recent experiments \cite{Ju_2015,Li_2016} show the gapless states localized along the DW \cite{Li_2016}, being robust conducting channels \cite{Ju_2015,Pelc_2015,San_Jose_2014,Koshino_2013}. 

%% FIGURE %%%
\begin{figure}[ht]
\centering
\includegraphics[width=\columnwidth]{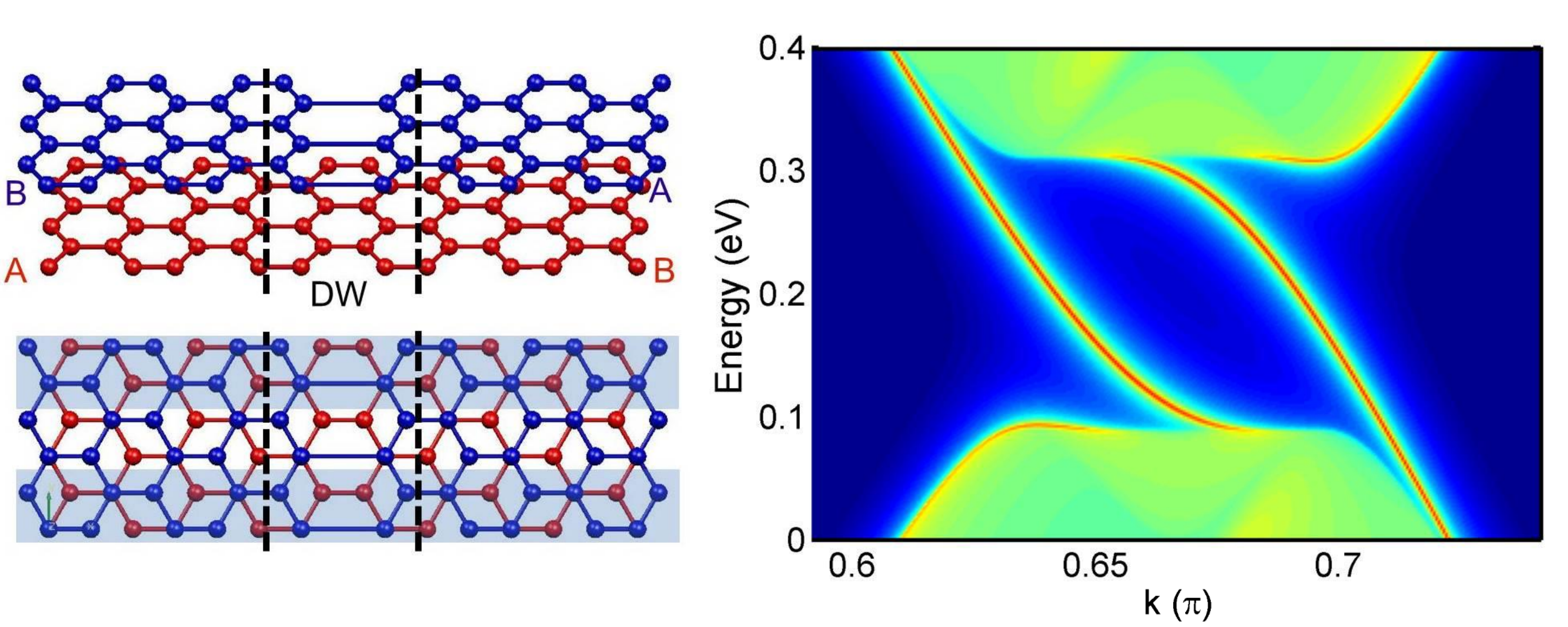}
\caption{\label{fig:sdw}
Left: Geometry of a minimal stacking domain wall in bilayer graphene, separating regions with AB ad BA stacking.
Right: Topological modes arising at the gap in gated bilayer graphene with a stacking domain wall. Lattice constant $a=1$.
}
 \end{figure}

Gapless states in bilayer graphene have also been studied theoretically in works focusing on their topological features \cite{Zhang_2013,Vaezi_2013,Martin_2008}. As long as the no-valley mixing condition is fulfilled, one can specify a topological quantity, namely, the valley Chern number \cite{Martin_2008,Zhang_2013}. Its change across the DWs accounts for the number of gapless states that appear under a gate potential. However, experiments and applications based on these conducting states need to clarify other important physical features, such as spatial localization, where using topology arguments would be complex. A microscopic model is better suited to identify the symmetry and localization of the states. A few relevant works on lattice models specifically study different stacking boundaries for a gate voltage close to the interlayer coupling \cite{Zhang_2013,Vaezi_2013}. However, a systematic analysis of the layer localization in these stacking DW states under varying gate voltages is still missing. 

In this work we show that the layer localization of gapless states in a gated BLG domain wall is tuned by an externally applied voltage. 
We demonstrate that at the domain wall the carriers are concentrated in the upper or bottom layer depending on the ratio between gate voltage and interlayer coupling. This dependence allows for switching the localization of topological states between layers by a change of the gate magnitude, implying that an additional degree of freedom, the layer, can play a role in BLG-based devices. Next, to explain the aforementioned behavior, we consider the symmetries of the bands and bonds formed between atoms in the gated system. We model a periodic array of stacking domain walls that, in contrast to the single DW, allows us to work with energy bands and wave functions. In this way, we identify the specific bands of the uncoupled layers which create the gapless states under interlayer coupling.

Controlling the layer localization by an external voltage opens a novel route for the manipulation of gapless states in electronic devices that could be denominated \textit{layertronics}, in analogy to valley- and spintronics. We propose that layer localization would be another tunable degree of freedom in BLG, in addition to valley and spin.

\section{Gated stacking domain wall: preliminary remarks}

We first review the main features of the system. The left panel of Fig. \ref{fig:sdw} shows a single domain wall between AB and BA regions in bilayer graphene joined in the zigzag direction. Strained or corrugated graphene presents larger and more realistic domain walls, but for our theoretical analysis we choose an abrupt boundary that allows for an AB/BA stacking change. Previous works \cite{Lin_2013,Vaezi_2013,Lin_2013,Ju_2015,Pelc_2015} have shown the robustness of topological states for smoother boundaries, with their main features preserved. For such a single boundary under a gate there are two topological gapless modes around the $K$ point, as shown in the right panel of Fig. \ref{fig:sdw}. The plot shows the local density of states (LDOS) resolved in energy $E$ and wave vector $k$. These calculations employ a Green's function matching method \footnote{The LDOS is calculated at the boundary in the zigzag direction. For details see Ref.\cite{Datta,Chico_1996,Jaskolski_2005,Santos_2009}} and a $p_z$ tight-binding model, with an intralayer hopping parameter $\gamma_0=-2.7$ eV and a single interlayer hopping $\gamma_1=0.1 \gamma_0$ \cite{Castro_2007,Ohta_2006}. There is another valley with another couple of gapless states with negative wave vectors with respect to Fig. \ref{fig:sdw} due to time-reversal symmetry. Note that the valley separation has motivated the proposal to employ such topologically protected modes for graphene valleytronics \cite{Rycerz_2007,Kundu_2016}.

\section{\label{sec:sdw_loc}Layer localization with gate voltage magnitude}

\begin{figure}[hb]
\includegraphics[width=\columnwidth]{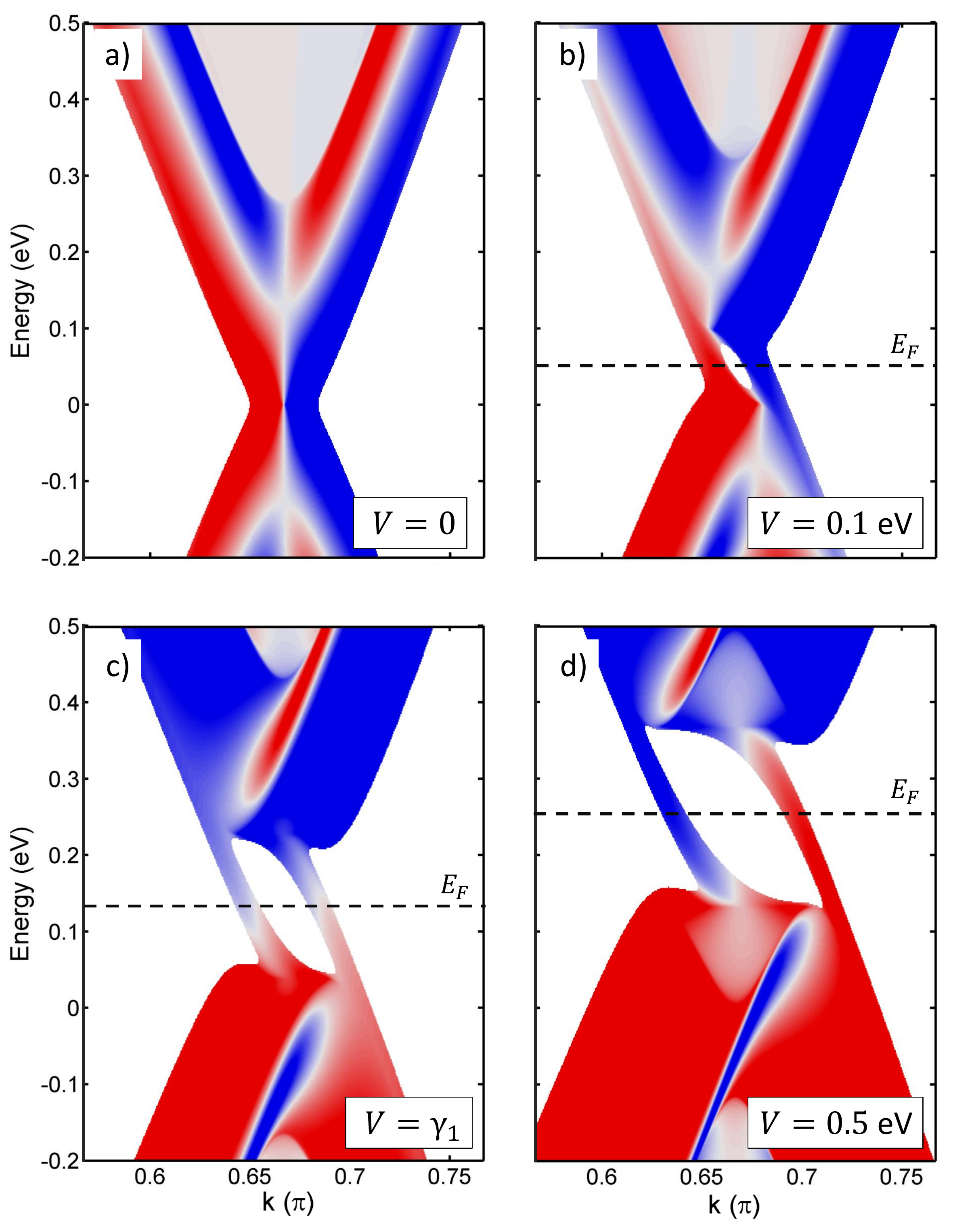}
\caption{\label{varyvg}
(Color online) Topological modes of bilayer graphene with a stacking domain wall: ungated (a) and when the gate voltage is applied to the bottom layer: $V=0.1$ eV (b), $V=\gamma_1$ (c) and $V=0.5$ eV (d). Color scale reflects the localization in top (blue) or bottom (red) layer. Dashed lines indicate the Fermi levels. For $V=0$ the Fermi level is at zero-energy, while for $V>0$ - in the middle of the gap, at $E=V/2$.
}
 \end{figure}

We report on the effect of layer localization exchange of topological states at a single stacking domain wall for small and large gate voltages. We use the same Green's function matching method and graphene model when there is an applied $V$. The LDOS around the $K$ valley projected in the boundary nodes is shown in Fig. \ref{varyvg} for different voltages. The localization in the top and bottom layers is presented by the color scale from blue to red \footnote{The LDOS ratio is not plotted, i.e. white areas outside the cones, when LDOS is lower than some small set value}. We start with the LDOS for ungated bilayer in panel (a). The LDOS localizes differently at the top and bottom layers because the symmetry between them is broken by the stacking domain wall. For a small gate voltage applied to the bottom layer, e.g. $V=0.1$ eV in panel (b), the gap states appear, as it is well known, but they turn out to be separated in the two layers. The state on the left side of the cone is more localized at the bottom layer, while the state on the right is at the top layer. This localization can be explained as a perturbation of the LDOS of the ungated bilayer for small voltages applied. For increasing voltages, around $V=\gamma_1$ in panel (c), the states become fully mixed between layers. Next, for a large voltage, e.g. $V=0.5$ eV in panel (d), the states are again separated in the top and bottom layer. This time, however, the state on the left side of the cone is more localized at the top layer, while the state on the right is at the bottom layer. Therefore, the topological states at the boundary atoms reverse localization for small and large voltages of the same sign. 

\begin{figure}[hb]
\includegraphics[width=\columnwidth]{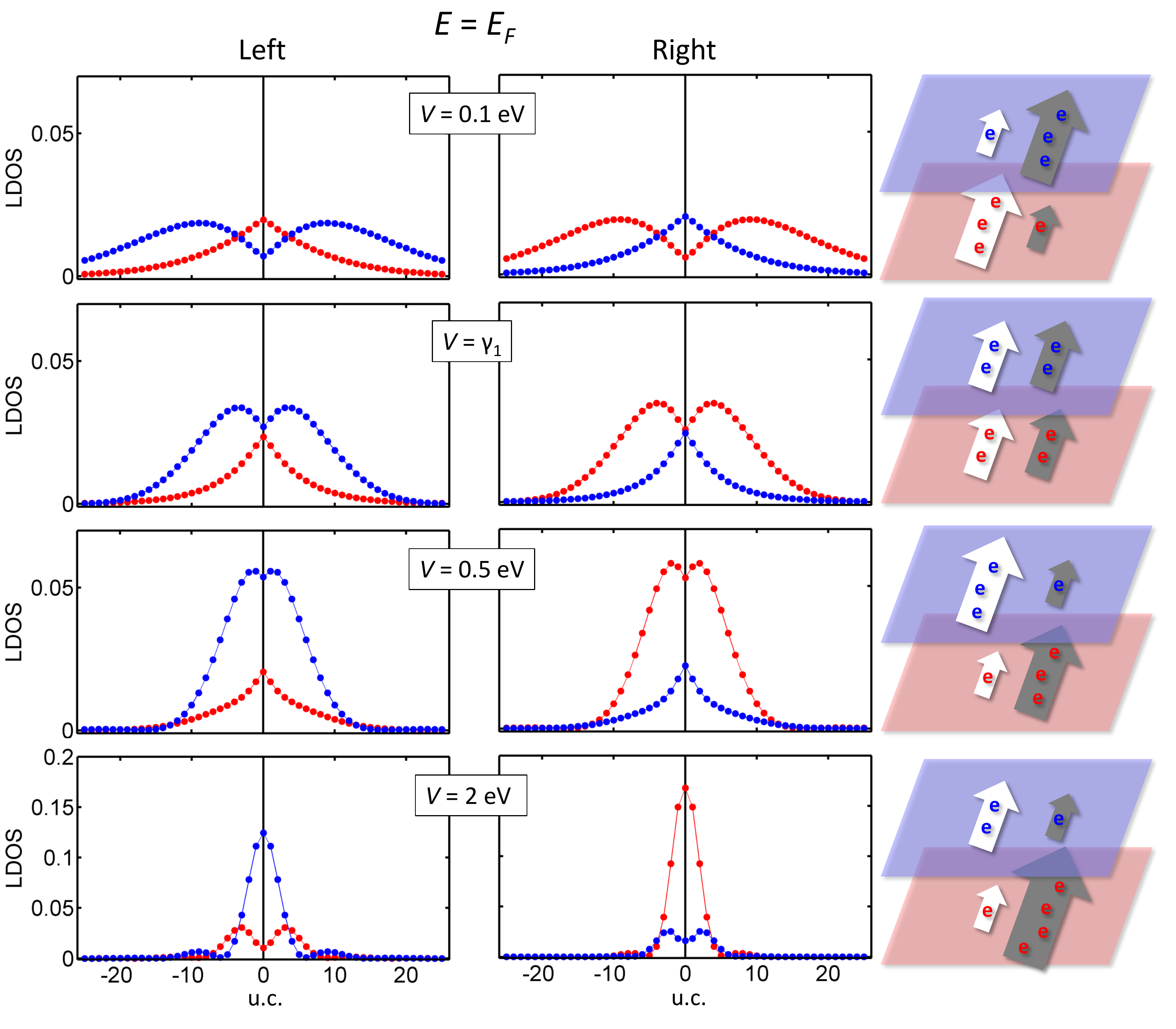}
\caption{\label{varyvg_2}
(Color online) Unit-cell averaged LDOS at the bottom (red) and the top layer (blue) plotted for both topologically protected states at the Fermi level for different gate voltages. The horizontal axis indicates the distance (in unit cells) from the boundary (DW). For each case we include schemes putting both states side by side - left and right topological states are presented by white and grey arrows, respectively. The sizes of the arrows and the number of electrons reflect the quantitative values of the localized LDOS per layer for each state. For high voltage the left and right topological states are not complementary, summing up to a large contribution to the bottom layer. 
}
 \end{figure} 
 
Now we explore how the localization of topological states extends away from the boundary atoms. We investigate their spatial distributions at the Fermi level, as shown in Fig. \ref{varyvg_2}. The LDOS of each state is averaged per rectangular zigzag unit cell (four atoms in each layer) and decomposed in the top and bottom layers. The spatial distributions show that localized states have their maxima close to the boundary region and decay in the adjacent unit cells \cite{Pelc_2015}. The weight of the LDOS at the top and bottom layers is interchanged by increasing the voltage, around $V\approx \gamma_1$, as commented above. 

It is also important to consider how far the layer interchange extends away from the boundary. We find that for small voltages, e.g $V=0.1$ eV, this LDOS swapping between layers extends for an experimentally relevant distance of about 4 nm. For $V=\gamma_1$ the LDOS of each state (left and right) is equally localized in the two layers just at the boundary, but it becomes layer-resolved up to distances of 40 unit cells, with left and right states having opposite layer localization. For larger potentials, e.g. $V=0.5$ eV, the LDOS of each state is clearly localized in one of the layers encompassing about 20 unit cells. For all these voltages we find that the LDOS of the left and right states at the Fermi level are symmetric under the interchange of layers, as seen in Fig. \ref{varyvg_2}. For much higher voltages, e.g. $V=2$ eV, the state localization in the bottom layer increases around the boundary. The states decay faster even when the gap remains almost constant (since $V>\gamma_1$). Surprisingly, the LDOS for the left and right states of the $K$ valley have different layer weight that produce an asymmetry between the two topological states. Note that for $V=0$, each valley is symmetric with respect to the Dirac point near the Fermi level. For high voltages, such symmetry is not present, giving rise to different, more dramatic layer localizations at these high voltages. It is surprising that the LDOS at the top and bottom layers are not complementary anymore (see bottom panels of Fig. \ref{varyvg_2}, $V=2$ eV). The LDOS localization in the $K$' valley is exactly the same as that for the $K$ valley. Taking into account both of them, we just have to double the weight of topological states. The lack of complementarity between such states at high voltages is crucial for engineering charge carriers and currents in distinct layers. 
 
 \begin{figure}[hb]
\includegraphics[width=\columnwidth]{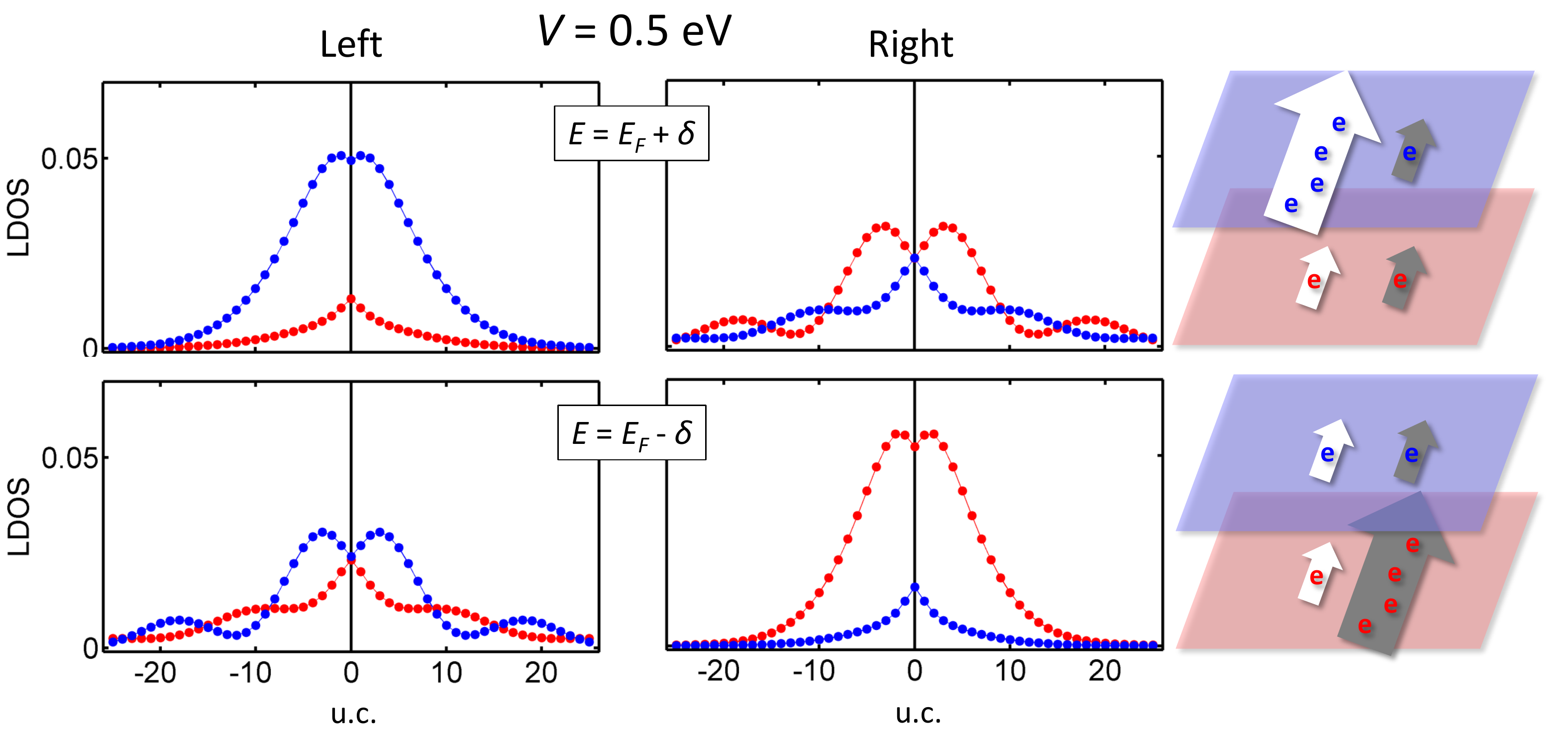}
\caption{\label{varyvg_3}
(Color online) Unit-cell averaged LDOS at the bottom (red) and the top layer (blue) plotted for both topologically protected states for $V=0.5$ eV above and below the Fermi level; $\delta=0.1$ eV. Schemes of the two topological states follow the notation given in Fig. \ref{varyvg_2}. The layer localization is now distinct at voltages accessible in experiments.
}
 \end{figure} 

Topological states in gated bilayer graphene run across the insulator gap, so that their behavior can also be investigated by varying the chemical potential. Note that up to now we have looked at the topological states at the Fermi energy. However, experiments are usually performed away from 
the neutrality point, due to doping or to the interaction with different substrates. Figure \ref{varyvg_3} shows the LDOS distributions of topological states above and below the Fermi level. The left and right topological states are no longer complementary. We find that the total charge is more localized on one of the layers, i.e., layer localization is tuned by doping. This effect seems similar to the above case with high voltages, but with significant differences because: (i) the topological states can be on either top or bottom layer, and (ii) the effect is available at lower gate voltage values. Additionally, the distribution of the topological states in the top or bottom layers can be controlled by the experimental gate voltage applied in doped samples, a finding that has to be taken into account for practical applications.

As far as we know, this effect based on the asymmetry in localization of topological states has not been noticed and not explored before. We believe that the reason for this omission is that for $V\approx \gamma_1$, as normally set in most calculations, the energy gap saturates at that value and the topological states are equally layer resolved \cite{McCann_2013,Vaezi_2013,Pelc_2015}. The layer distribution for these potential values is fully mixed at the boundary atoms. 
  
\section{Periodic stacking domain walls}

\begin{figure}[hb]
\centering
\includegraphics[width=\columnwidth]{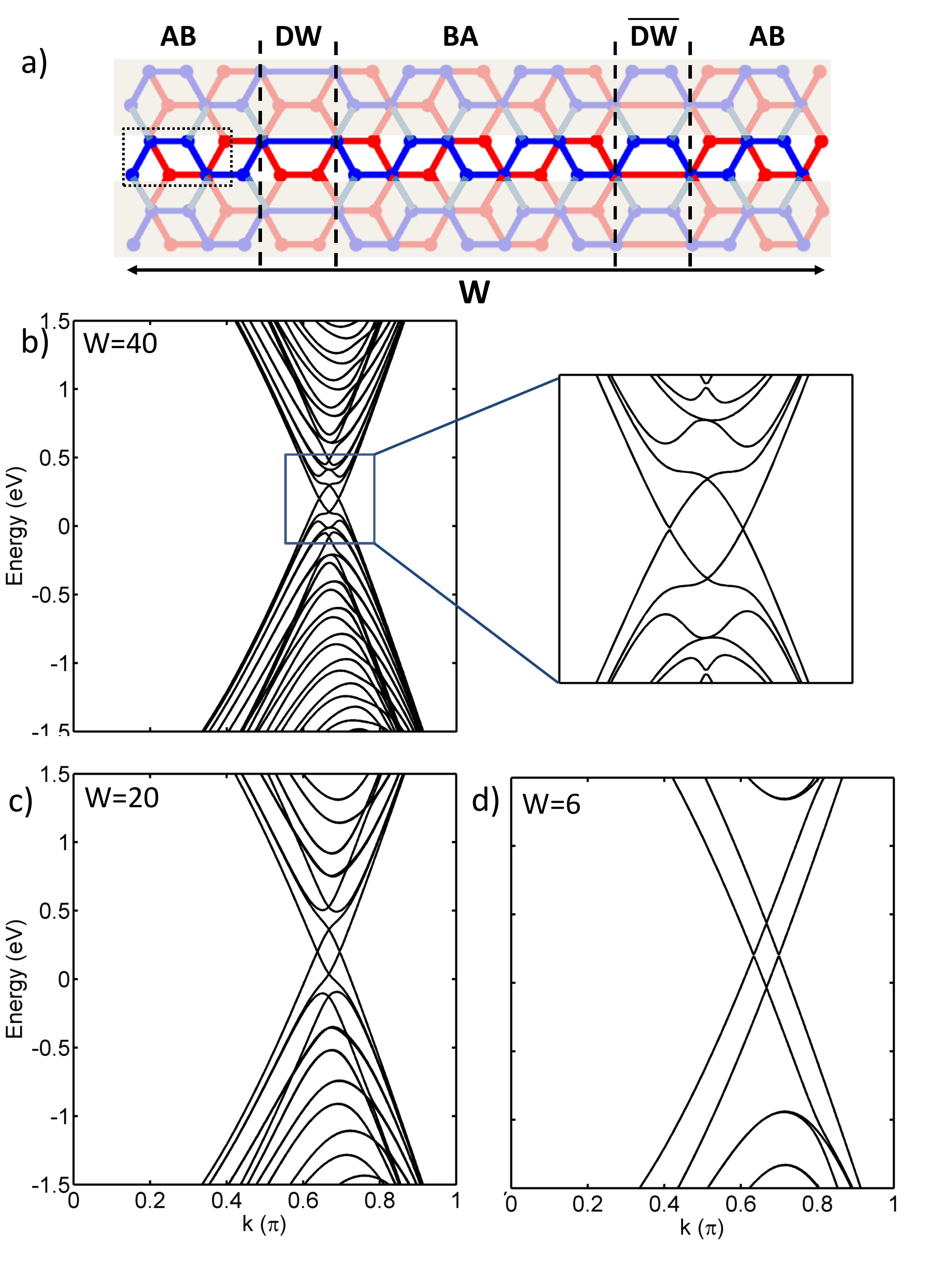}
\caption{\label{fig:sldw}
(Color online) (a) Schematic geometry of the unit cell for a bilayer superlattice with two stacking domain walls. AB, BA and domain walls are delimited with dashed lines. The 8-atom cell used as unit of length is marked with a dotted rectangle. (b) Band structure near the $K$ point along the $k$ zigzag direction of a $W=40$ BS-DW. A zoom of the topological modes is shown at the right panel. Band structures for (c) $W=20$ and (d) $W=6$ BS-DW in the same $k$ direction as (b). In all cases the gate voltage is set to $V=0.4$ eV. 
}
\end{figure}

In order to gain physical insight into the remarkable variation of the spatial distribution of these modes, we need to examine the symmetry of the corresponding wavefunctions. To this end, we study periodic systems, i.e. bilayer superlattices of stacking domain walls along the zig-zag direction \cite{Jaskolski_2005,Santos_2009}, which we label BS-DW \footnote{The choice of periodic systems allows us to avoid the interplay between topological states and edge states characteristic for zigzag graphene edge \cite{Jaskolski_2011}}. The length $W$ of the bilayer superlattice unit cell is measured by the number of 8-atom units, as marked in Fig. \ref{fig:sldw} (a). The system is made periodic perpendicular to DW by including two stacking boundaries with reverse effect: one to change from AB to BA (marked as DW) and another one to change from BA to AB stacking ($\overline{\rm DW}$), as shown in Fig. \ref{fig:sldw} (a). Strain is accumulated in the bonds of the top layer in one case and of the bottom layer in the second 
\footnote{In fact, the boundaries consist of a different bonding of the atoms of the bottom and the top layer, without actually modifying the numerical value of the hopping.}. 

Figure \ref{fig:sldw} (b) presents the band structure of the superlattice for a large unit cell, $W=40$. The dispersion relation is plotted for the wave vector $k$ along the stacking boundary direction, in order to compare to the $k$-resolved LDOS presented in Fig. \ref{fig:sdw}. The spectra are in fact rather similar, with the obvious difference that in the superlattice there are now four topological modes (two pairs that cross each other) due to the existence of two stacking domain walls in the unit cell, one pair with positive velocity and the other with opposite slope. The dispersion of these modes is clearly seen in the zoom. Decreasing the superlattice spatial period $W$, as in Fig. \ref{fig:sldw}(c,d), the linear portions of the topological bands become larger. So the peculiar shape of topological bands as presented in the single domain wall of Fig. \ref{fig:sdw}, or in the zoom for a large superlattice, as in \ref{fig:sldw}(b), is due to the repulsion from the rest of the valence and conduction bands, not being an intrinsic property of these modes. Note that these modes disappear for $V=0$ or if the interlayer hopping is eliminated, because either way the gap is closed, as mentioned before. One common procedure to study the occurrence of gapless modes is to consider the gate potential as a perturbation.

It is noteworthy that for $V>0$ and $\gamma_1=0$ the bands of the constituent layers cross near the Fermi level. For nonzero $\gamma_1$ the crossing bands interact and split yielding the energy gap in the case of pristine bilayer. However, when a stacking domain wall is imposed, two bands still persist in the gap. The treatment of $\gamma_1$ as a perturbation allows to recognize the bands of single graphene layers that give rise to the presence of topological bands in the energy gap. 

\begin{figure}[hb]
\centering
\includegraphics[width=\columnwidth]{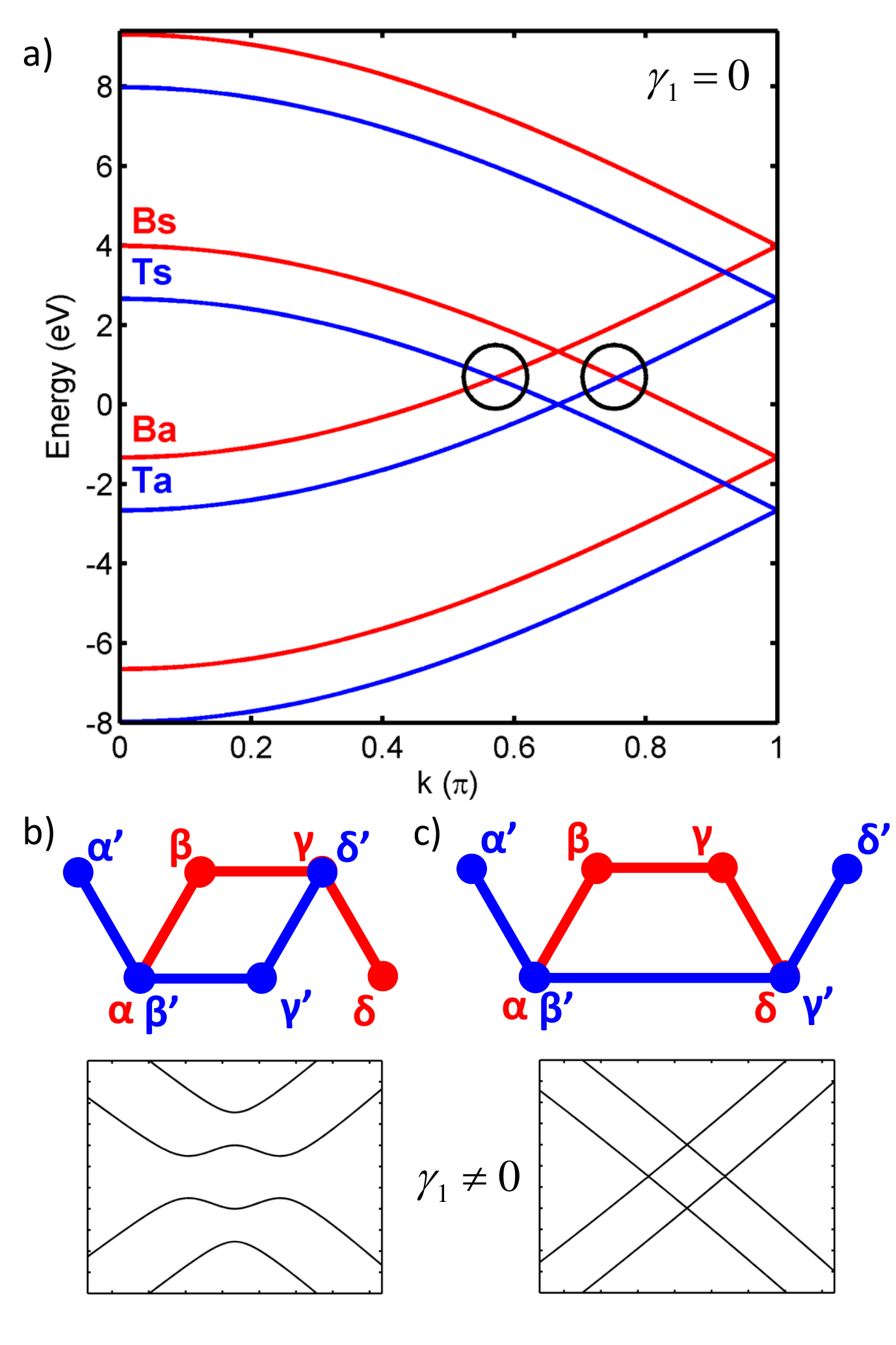}
\caption{\label{fig:uncoupled}
(Color online) 
(a) Bands of two uncoupled layers of graphene, with a gate voltage applied to the bottom layer, for an 8-atom rectangular unit cell, shown below. 
Red and blue bands correspond to bottom and top layers, respectively. Bands are labeled according to localization and symmetry; circles mark the crossing points analyzed below, see text. Rectangular 8-atom unit cells, with all atoms labeled, for (b) AB bilayer graphene and (c) minimal double stacking domain wall, i.e., a BS-DW with $W=1$. The resulting band structures close to the crossing points are shown below the respective unit cells. Greek symbols in (b) and (c) enumerate the nodes in the 8-atom unit cell.
}
\end{figure}

\section{Insight into topological gapless states}

\subsection{Gaps and band crossing points near the Fermi level}

We consider two uncoupled graphene layers with a gate potential $V$ applied to the bottom layer and we switch on the interlayer hopping to study its role in the appearance of the topologically protected bands. 
Without hopping, the energy structure of two pristine graphene layers with a voltage difference applied is gapless, as shown in Fig. \ref{fig:uncoupled} (a). We have chosen an 8-atom unit cell to compare more easily to the band structures of stacking DW superlattices. The bands of the gated bottom layer are shifted in energy by an amount $V$ with respect to the top layer bands. At $k=2/3\pi$ the pair that belongs to the top ungated layer crosses at $E=0$, while the pair of bands of the bottom gated layer cross at $E=V$. Note that the bands originating from different layers cross at $E=\frac{1}{2}V$. One pair crosses for $k \lesssim 2/3 \pi$ and another pair for $k \gtrsim 2/3 \pi$. The bands are labeled $B$, $T$, indicating that they belong to the bottom and top layers, respectively, and $s$, $a$, due to the symmetric or antisymmetric character of the corresponding wavefunctions. 

When the interlayer interaction is switched on in the BLG case (Fig. \ref{fig:uncoupled} (b)), the crossing bands mix and split because they form pairs of bonding and antibonding states and a gap opens. The resulting band structure near the Fermi level has the well-known Mexican hat shape (see the bands plotted in Fig. \ref{fig:uncoupled} (b) below the unit cell) \cite{Castro_2007}. However, when the nodes are connected producing the stacking defects, as in Fig. \ref{fig:uncoupled} (c), a pair of states remains in the gap, shown in the band structure depicted under the corresponding unit cell. The resulting structure is a BS-DW with $W=1$. 
A key question that we address below is why these states are gapless and how they arise from the bands of pristine graphene. 
 
\begin{figure}[hb]
\centering
\includegraphics[width=\columnwidth]{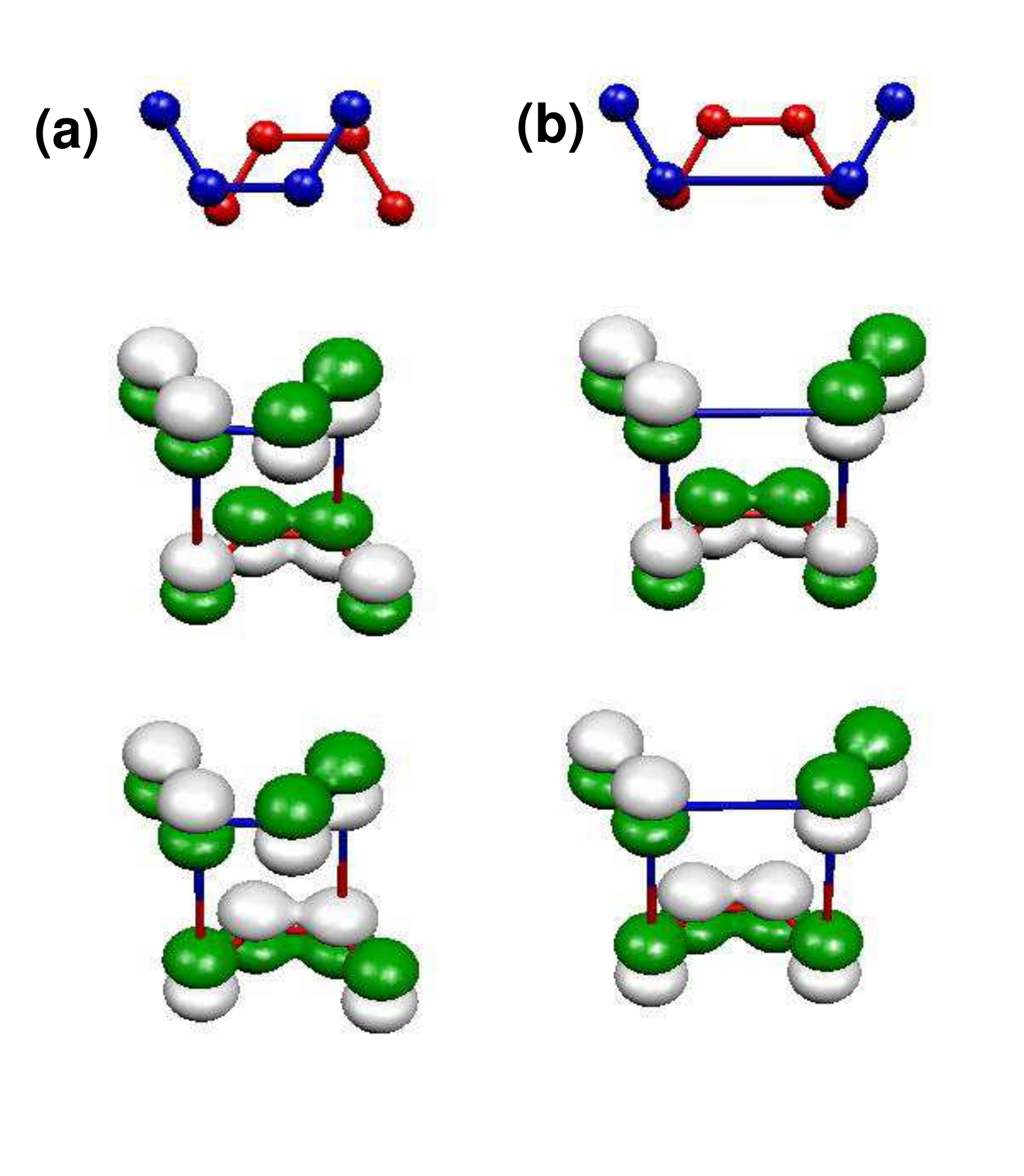}
\caption{\label{fig:bonds}
(Color online) Scheme of the interlayer bonds between $p_z$ orbitals at the left crossing point (see Fig. \ref{fig:uncoupled} (a)), for (a) bilayer graphene with an 8-atom unit cell and (b) a BS-DW of width $W=1$. Top and bottom layers are colored in blue and red, respectively. The bonding and antibonding orbitals between the $Ts$ and $Ba$ wavefunctions are shown below. Green and white colors denote the sign of $p_z$ lobes. The vertical blue / red lines show the atoms in the top layer that couple to atoms in the bottom layer. For pristine bilayer graphene, the combinations of the $p_z$ orbitals yield different bonding and antibonding solutions, $\Psi_{\pm}$. No such solutions can be formed in the BS-DW case.
}
 \end{figure}

We focus on the left crossing point marked with a circle in Fig. \ref{fig:uncoupled} (a). The two bands crossing therein are labeled $Ts$ and $Ba$, due to their symmetry and localization. The bonding and antibonding $p_z$ orbital combinations formed due to the interlayer interaction are schematically illustrated in Fig. \ref{fig:bonds} (a). However, for a domain wall, there are always topological modes crossing the gap. Figure \ref{fig:bonds} (b) graphically shows that it happens because it is not possible to form bonding and antibonding combinations of the top and bottom layer wave functions simultaneously for both pairs of the overlapping nodes from different layers in a DW. 

\subsection{Basis functions for the crossing bands}

We next treat the interlayer hopping as a perturbation to the bands of the uncoupled layers. The starting basis is given in terms of the uncoupled bands depicted in Fig. \ref{fig:uncoupled} (a). For the 8-atom unit cell employed therein, and explicitly labeled in Fig. \ref{fig:uncoupled} (b), 
we have

\begin{equation}
\label{eq:bas4}
\begin{array}{l}
\Psi_{Ba,k}=\frac{1}{2}(\phi_\alpha+ e^{ikd} \phi_\beta -e^{ikd} \phi_\gamma - \phi_\delta)\\
\Psi_{Bs,k}=\frac{1}{2}( \phi_\alpha - e^{ikd}\phi_\beta -e^{ikd} \phi_\gamma + \phi_\delta)\\
\Psi_{Ta,k}=\frac{1}{2}(e^{ikd}\phi_{\alpha'} + \phi_{\beta'} -\phi_{\gamma'} - e^{ikd}\phi_{\delta'}) \\
\Psi_{Ts,k}=\frac{1}{2}( e^{ikd}\phi_{\alpha'} - \phi_{\beta'} -\phi_{\gamma'} + e^{ikd} \phi_{\delta'})
\end{array} 
\end{equation}
where the labels refer to their layer localization and their symmetry, as in Fig. \ref{fig:uncoupled} (a); $d$ is the distance between contiguous rows of atoms in the zigzag direction (see Fig. \ref{fig:bonds}); $\phi_\mu$ denotes the $p_z$ orbital in the $\mu$ atom, with $\mu$ running from $\alpha$ to $\delta$ in the bottom layer and the same labels with primes in the top layer (see Fig. \ref{fig:uncoupled} (b) and (c)). The wave vector $k$ dependence is explicitly indicated as a subscript, but we omit it from now on for the sake of simplicity.

For connected layers, the degeneracy at the crossing points of the uncoupled system ($k \sim \frac{2}{3}\pi$), marked in Fig. \ref{fig:uncoupled} (a) with circles, can be lifted. 
Focusing in the band wavefunctions at the left crossing point, we construct the bonding and antibonding combinations 
$\Psi_{+}=\frac{1}{\sqrt{2}}(\Psi_{Ts} + \Psi_{Ba})$
and 
$\Psi_{-}=\frac{1}{\sqrt{2}}(\Psi_{Ts} - \Psi_{Ba})$, respectively. 
When the interlayer coupling is switched on, these states are shifted in energy by an amount given by $\braket{ \Psi_{\pm} | H^{TB} | \Psi_{\pm} }$, where $H^{TB}$ is the interlayer coupling Hamiltonian. 

By connecting the layers, pristine gated bilayer graphene is obtained; to this purpose the connected atom pairs are $\alpha-\beta'$ and $\gamma-\delta'$, see Fig. \ref{fig:uncoupled} (b). The energy shift of the bonding state from the energy at the crossing point is given by $$\braket{ \Psi_{+} | H^{TB}_{BLG} | \Psi_{+} }=-\frac{1}{2}|\gamma_1|.$$ Analogously, the energy shift of the antibonding state is $$\braket{ \Psi_{-} | H^{TB} _{BLG}| \Psi_{-} }= \frac{1}{2}|\gamma_1|, $$ so that the total gap equals $|\gamma_1|$. This is why the bands $Ts$ and $Ba$ split at the left crossing point of Fig. \ref{fig:uncoupled} (a). At the right crossing point it happens the same, but now for the $Ta$ and $Bs$ bands. 

Along the same line of reasoning, we discuss why two pairs of the overlapping bands survive in a periodic gated domain wall. The unit cell is shown in Fig. \ref{fig:uncoupled} (c). In the previous case the two pairs of the overlapping nodes from the red and blue layers were different. At the left crossing point, the two $\Psi_{Ts}\pm \Psi_{Ba}$ functions yield a bonding combination of the $p_z$ orbitals at one pair of the connected nodes, but an antibonding combination at another pair of nodes, see Fig. \ref{fig:bonds}. In other words, the interlayer Hamiltonian changes, because now the connected nodes are $\alpha-\beta'$ and $\delta-\gamma'$, as depicted in Fig. \ref{fig:uncoupled} (b). With this different coupling, $$\braket{ \Psi_{\pm} | H^{TB}_{DW} | \Psi_{\pm} }=0, $$ and the bands $Ts$ and $Ba$ still have a crossing point. 
A similar analysis can be performed for the right crossing point for the $Ta$ and $Bs$ bands. As a result, we end up with two pairs of states connecting the valence and conduction band, i.e., the topological modes. 

\begin{figure}[hb]
\centering
\includegraphics[width=\columnwidth]{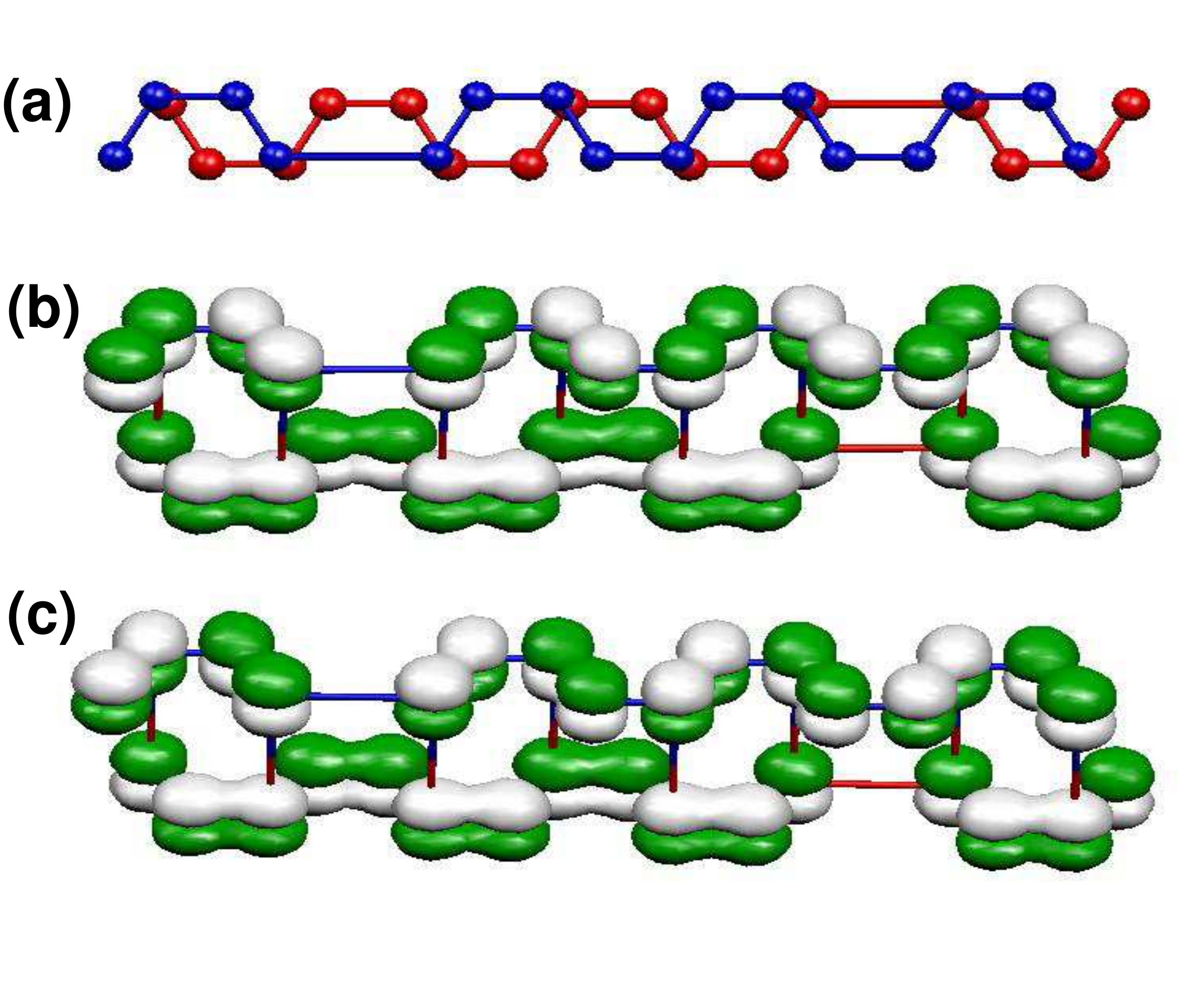}
\caption{\label{fig:fig6}
(Color online) 
Geometry (a) and distribution of the signs of coefficients of the wave functions $\Psi_{+}$ (b) and $\Psi_{-}$ (c) in the unit cell of BS-DW with $W=4$. }
 \end{figure}

This reasoning can be applied to a BS-DW of arbitrary width. In such a case, one can choose a bilayer graphene superlattice of the same width 
$W=n$, consider the uncoupled layer case, which yields a band structure similar to Fig. \ref{fig:fig6} (a) but with $4n$ bands. In such instance, the bands crossing at $E_F$, i.e., the $2n$ and $2n+1$ bands have exactly the same antisymmetric and symmetric character as those at the crossing points analyzed for the 8-atom case. 
The difference in the wavefunctions is the normalization factor, being $\frac{1}{\sqrt{N}}$, with $N=4n$ is the number of atoms in the unit cell. When the coupling is switched on and the geometry corresponds to a BS-DW, the energy shifts corresponding to $\Psi_{+}$ and $\Psi_{-}$ are zero, so there are two pairs of bands crossing the gap. 
Figure \ref{fig:fig6} (b) and (c) illustrates the $\Psi_{\pm}$ band wave functions for the case of $W=4$. The sign of the $p_z$ orbitals at the domain walls are exactly as those found for the minimal BS-DW. 

\begin{figure}[hb]
\centering
\includegraphics[width=\columnwidth]{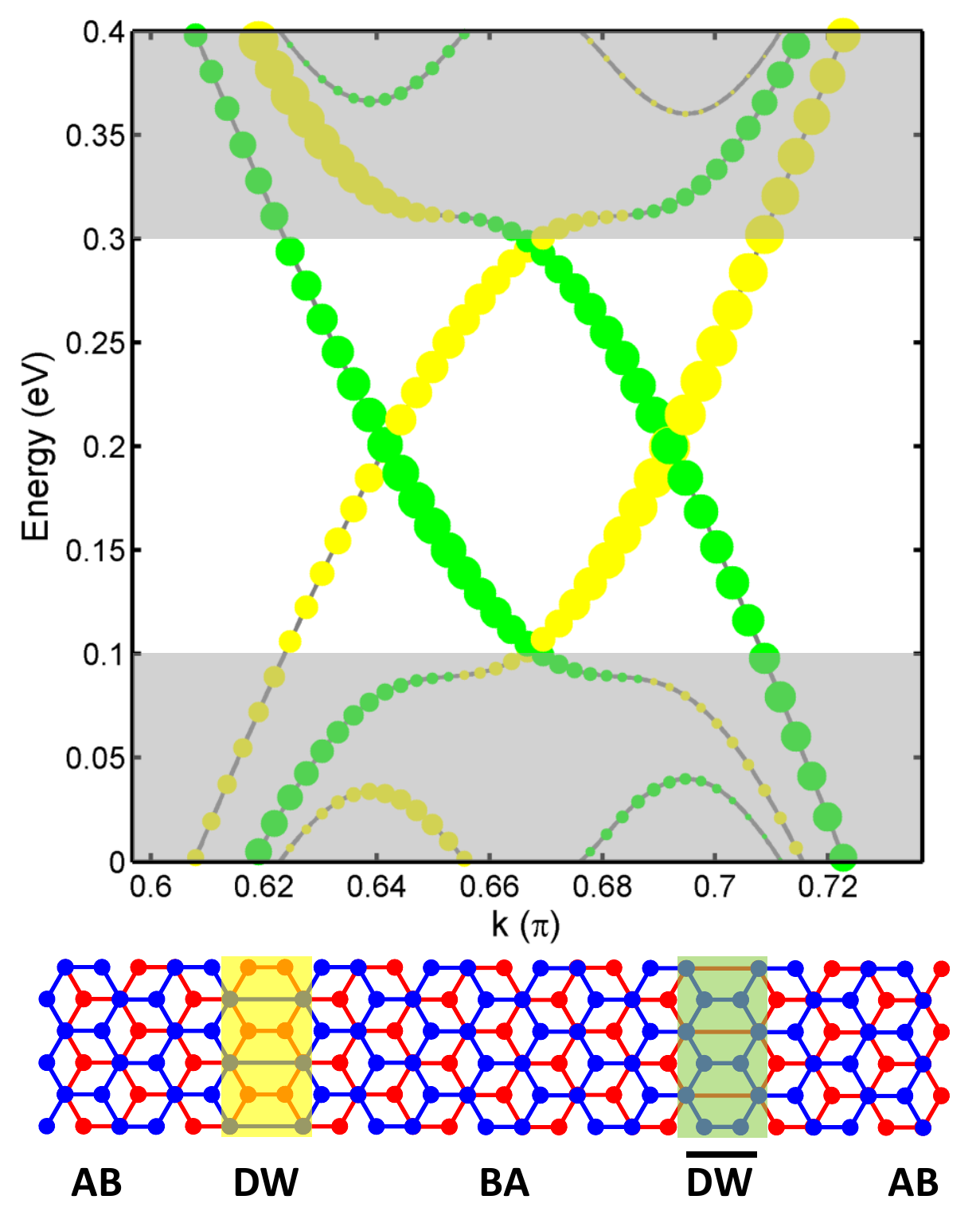}
\caption{\label{fig:fig7}
(Color online) 
Bands of BS-DW of width $W=40$ at gate voltage $V=0.4$ eV resolved into the two stacking domain walls present in unit cell: DW - yellow and $\overline{\rm DW}$ - green, marked also on the scheme below. The size of the dots is proportional to the module of wavefunction on the particular stacking domain wall. Shaded areas mark the valence and conduction band continua for the single DW case. Four localized gap bands are in the non-shaded area.
}
 \end{figure}

\begin{figure}[hb]
\centering
\includegraphics[width=\columnwidth]{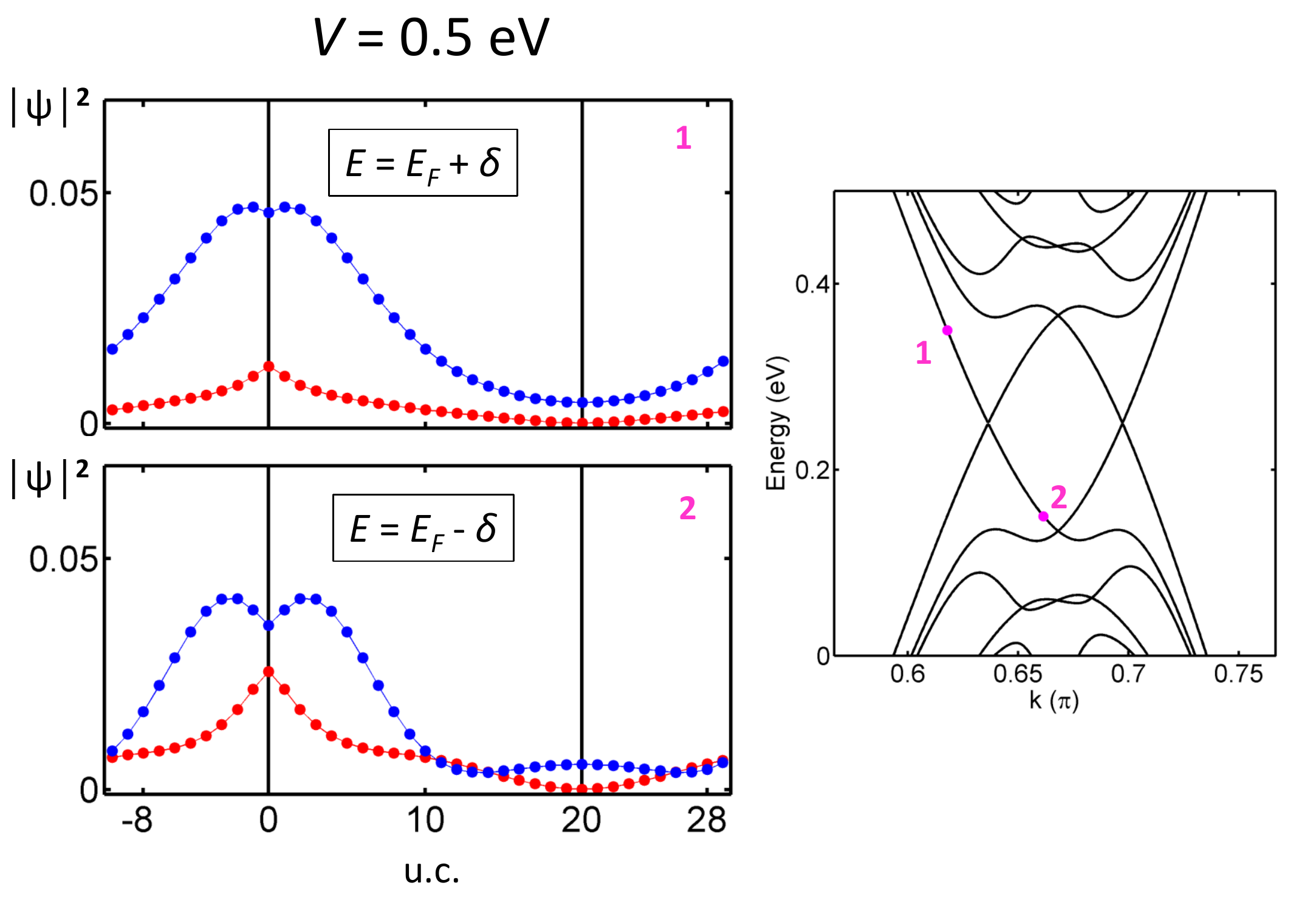}
\caption{\label{fig:fig_wf}
(Color online) 
(Color online) Unit-cell averaged $\Psi^*\Psi$ of BS-DW of width $W=40$ at the bottom (red) and the top layer (blue) plotted for one of the topologically protected states for $V=0.5$ eV above and below Fermi level; $\delta=0.1$ eV. The difference in localization is similar to a single domain wall in Fig. \ref{varyvg_3}.}
 \end{figure}

As commented before, in a BS-DW one pair of the bands crossing the gap has positive velocity and the other pair has a negative slope. In the smallest BS-DW with $W=1$, the states at the boundaries are unavoidably mixed. For large $W$, each pair belongs to a different stacking domain wall in the unit cell, being spatially separated, as illustrated in Fig. \ref{fig:fig7}. The bands for $W=40$ are resolved in the two stacking boundaries, DW and $\overline{\rm DW}$. When one stacking DW is present in the system, one pair of bands is in the gap with the same velocity, as shown in Fig. \ref{fig:sdw}. The sign of the velocity is related to the change of stacking, either AB to BA or vice versa. Further, the different localization between layers predicted from the periodic calculations are compared with the case for a single DW, as shown in Sect. \ref{sec:sdw_loc}. In Fig. \ref{fig:fig_wf} we present the wavefunctions corresponding to $E_f\pm \delta$ around the DW. We find that a DW either isolated or in a periodic arrangement behaves the same way.

\subsection{Continuum Hamiltonian for topological states}

We generalize the above discussion to $k$ values away from the band crossing points. The Hamiltonian $H$ of each layer close to the Dirac point is represented in the basis (\ref{eq:bas4}), given by $\Psi_{Ba}$ and $\Psi_{Bs}$ for one layer, and $\Psi_{Ta}$ and $\Psi_{Ts}$ for the other 
layer. For a monolayer, $H$ is a $2\times2$ diagonal matrix with linear $k$, and $-k$ values in the diagonal elements. For gated bilayer graphene 
we have to double this matrix \cite{McCann_2013,Castro_2009}. Using the basis $\{\Psi_{Ba},\Psi_{Bs},\Psi_{Ta},\Psi_{Ts}\}$ we obtain the interaction Hamiltonian between layers that introduces the hopping $\gamma_1$ in the non-diagonal elements. Employing the expression of these basis vectors in terms of the localized orbitals $\phi_{\mu}$ given in 
Eq. (\ref{eq:bas4}) and the geometry of the interlayer bonds shown in Fig. \ref{fig:uncoupled} (b), the Hamiltonian for bilayer graphene is written 

\begin{equation}
\label{eq:h1}
H_{BS}=\left( \begin{array}{cccc}
\alpha k + V & 0 & \frac{1}{2}\gamma_1 &- \frac{1}{2}\gamma_1 \\
0 & -\alpha k + V & \frac{1}{2}\gamma_1 & -\frac{1}{2}\gamma_1 \\
 \frac{1}{2}\gamma_1 & \frac{1}{2}\gamma_1 & \alpha k &0 \\
-\frac{1}{2}\gamma_1 & -\frac{1}{2}\gamma_1 &0 & -\alpha k 
\end{array}\right),
\end{equation}
where $\alpha=\sqrt{3}a\gamma_0 /2\hbar$, and $a$ is the graphene lattice constant. For $V=0$ the eigenvalues of this Hamiltonian have a quadratic dependence on $k$, as it should be in bilayer graphene \cite{McCann_2013}. Two of them are degenerate at the Dirac point. For $V\ne 0$ the gap opens and the low energy bands versus $k$ show the Mexican hat shape. In the gated system the energy gap $E_g$ depends on the parameters $V$ and $\gamma_1$. For $V < \gamma_1$, $E_g \approx V$, while for $\gamma_1 < V$, $E_g \approx \gamma_1$. The Hamiltonian (\ref{eq:h1}) gives an appropriate description of the low-energy bands of BLG.

In the same way, i.e., using the basis vectors given in (\ref{eq:bas4}), we derive a Hamiltonian for bilayer graphene with periodic stacking domain walls as, 
\begin{equation}
\label{eq:h2}
H_{BS{\rm-}DW}=\left( \begin{array}{cccc}
\alpha k + V & 0 & \frac{1}{2}\gamma_1 & 0 \\
0 & -\alpha k + V & 0 & -\frac{1}{2}\gamma_1 \\
 \frac{1}{2}\gamma_1 & 0 &\alpha k &0 \\
0 & -\frac{1}{2}\gamma_1 &0 &-\alpha k 
\end{array}\right).
\end{equation}
The Hamiltonian (\ref{eq:h2}) includes a $k$ dependence, but it does not mix $k$ and $-k$ values inside or between the layers. The eigenvalues remain linear in $k$, so for $V \ne 0$ they cross the energy gap constituting the topologically protected gap states. Its eigenvalues have the following analytical expression:
\begin{equation}
E = \pm \alpha k +V/2\pm \frac{1}{2} \sqrt{ \gamma_1^2+V^2}.
\end{equation}
The corresponding eigenvector components for velocity $+k$ are
\begin{equation}
(0,\frac{-V\pm \sqrt{ \gamma_1^2+V^2}}{ \gamma_1},0,1)
\end{equation}
and for velocity $-k$
\begin{equation}
 (\frac{ V\pm \sqrt{ \gamma_1^2+V^2}}{ \gamma_1}, 0,1,0).
\end{equation}

Note that the localization of the states will behave differently in the two gate voltage regimes. In the limit of $V \ll \gamma_1$ the eigenvectors are mixed between layers - in contrast to the single domain wall case, where we observe localization reversal at the boundary. The reason for layer separation was the asymmetry between layers, which is no longer present in BS-DW.

In the limit of $V \gg \gamma_1$ the eigenvectors components of the Hamiltonian given by Eq. \ref{eq:h2} approach to $(0,0,0,1)$ and $(0,V/\gamma_1,0,1)$, and to $(0,0,1,0)$ and $(V/\gamma_1,0,1,0)$. This means that each pair of the gap states with the same velocity (+$k$ or $-k$) has the corresponding wave functions localized in different layers. The wave functions in each pair have the same symmetry, i.e., $a$ or $s$ (see Eq. \ref{eq:bas4}). Therefore, the topologically protected gap states look like the single layer bands that cross at the Fermi level. This continuum Hamiltonian with four bands allows us to predict the relevant properties of topological states in stacking domain walls.

%%%%%%%%%%%%%%%%%%%%%%%%%%%%%%%%%%%%%%%%%%%%%%%%%%%%%%%%%%%%%%%%%%%%%%%%%%%
%%% SUMMARY %%%
%%%%%%%%%%%%%%%%%%%%%%%%%%%%%%%%%%%%%%%%%%%%%%%%%%%%%%%%%%%%%%%%%%%%%%%%%%%%
\section{\label{sec:conc} Discussion and conclusions}

We have investigated the gapless states with topological character that appear in gated bilayer graphene with stacking domain walls. By employing atomistic models, we find that each of the two topological states in a valley is layer-resolved; furthermore, their localization is switched between the top and bottom layer by varying the magnitude, but not the sign, of the gate voltage. Therefore, besides the valley and sublattice degrees of freedom, these states can also be labeled by a layer index. 

Our findings also bring forth an atomistic understanding of the origin of gapless states with topological character. We have analyzed the chemical bond formation between gated layers of graphene with domain walls 
to elucidate the appearance of gapless states. We have also provided a continuum model for the gapless states that correctly describes the swithing in layer localization of the topologically protected states. For a large gate voltage, i.e., above the interlayer coupling, the layer localization presents the standard trend between top and bottom layer for a particular voltage orientation. However, for a small gate voltage, the asymmetry between layers introduced by the stacking domain walls prevails, 
so the gapless states are slightly perturbed by this voltage, having an opposite layer localization. 

Furthermore, we have shown that the layer spatial distribution of the topologically protected states is modified by doping and tuned by the gate voltage. The layer LDOS can be directly measured with an STM; this tool can also allow for further engineering on topologically protected states by adsorbing and moving molecules along the stacking domain walls. 
Controlling the carriers localization in distinct layers along domain walls would open the possibility for the design of {\em layertronic} devices, which could be exploited in addition to other degrees of freedom, like valley, spin and charge, in graphene-based electronics.


\begin{acknowledgements}
This work was partially supported by Projects FIS2015-64654-P and FIS2016-76617-P of the Spanish Ministry of Economy and Competitiveness MINECO, the Basque Government under the ELKARTEK project (SUPER), and the University of the Basque Country (Grant No. IT-756-13). L. C. acknowledges the hospitality of the Donostia International Physics Center.
\end{acknowledgements}

%\bibliography{bib_sdw}

\end{document}